\documentclass[twocolumn,prb,showpacs,aps,amsmath]{revtex4}

\usepackage{graphicx}
\usepackage[dvips]{color}
\usepackage{epsfig,euscript}

\begin{document}
\title{Hole-depletion of ladders in Sr$_{14}$Cu$_{24}$O$_{41}$ induced by
  correlation effects}

\author{V. Ilakovac$^{1,2,*}$, C. Gougoussis$^{3}$, M. Calandra$^{3}$, 
N. B. Brookes$^{4}$, V. Bisogni$^{4}$, S. G. Chiuzbaian$^{1}$,\\
J. Akimitsu$^{5}$, O. Milat$^{6}$, S. Tomi\'c$^{6}$, and C. F. Hague$^{1}$}

\email{vita.ilakovac-casses@upmc.fr}

\affiliation{$^1$Laboratoire de Chimie Physique Mati\`ere et Rayonnement, UPMC, CNRS, F-75231 Paris, France\\
$^2$Universit\'e de Cergy-Pontoise, F-95031 Cergy-Pontoise, France\\
$^3$Institut de Min\'eralogie et de Physique des Millieux Condens\'ees, UPMC, CNRS, F-75252, Paris, France\\
$^4$ESRF, B.P. 220, 38043 Grenoble Cedex, France\\
$^5$Departement of Physics, Aoyama-Gakuin University, Setagaya, Tokyo 157-8572, Japan\\
$^6$Institut za fiziku, P.O. Box, HR-10001 Zagreb, Croatia
}

\date{\today}

\begin{abstract}
The hole distribution in Sr$_{14}$Cu$_{24}$O$_{41}$ is studied by low
temperature polarization dependent 
O K Near-Edge X-ray Absorption Fine Structure  
measurements and state of the art electronic structure calculations
that include core-hole and correlation effects in a
mean-field approach.  
Contrary to all previous analysis, based on semi-empirical models,  
we show that correlations and antiferromagnetic ordering favor the 
strong chain hole-attraction. 
For the remaining small number of holes accommodated on ladders, 
leg-sites are preferred to rung-sites. 
The small hole affinity of rung-sites explains naturally
the 1D - 2D cross-over in the phase diagram of (La,Y,Sr,Ca)$_{14}$Cu$_{24}$O$_{41}$.
\end{abstract}


\maketitle

\section{\label{sec:Intro}Introduction}
\begin{figure}[t]
\begin{center}
\includegraphics [width=7.5cm,angle=0]{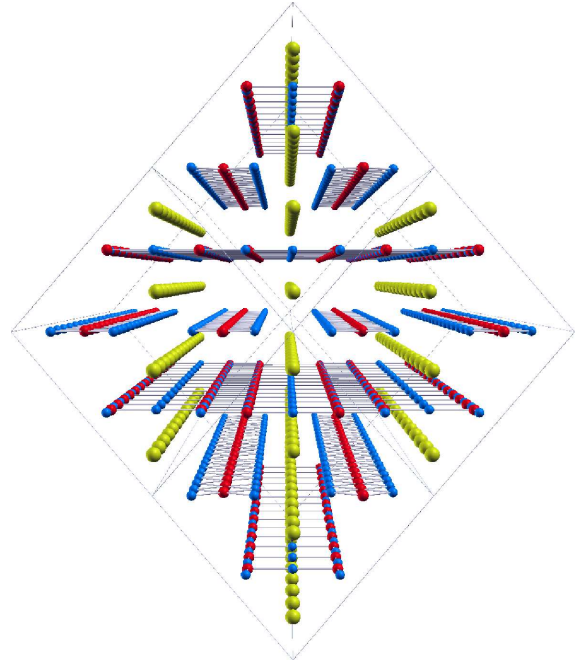}
\caption{\label{fig1} (Color online) 
3D view along the \textbf{c} crystal axis of four 316-atom AF unit cells of Sr$_{14}$Cu$_{24}$O$_{41}$. 
The structure is composed of alternately stacked $chain$ layers, 
constituted of CuO$_2$ linear edge-sharing CuO$_{4}$ squares, and $ladder$ planes, 
made up of two zig-zag strings of Cu$_2$O$_3$ corner-sharing squares. 
They are parallel to the (\textbf{a},\textbf{c})-plane and separated by Sr atoms. 
Cu/O/Sr atoms are colored in red/blue/yellow.}
\end{center}
\end{figure}

The quasi-one-dimensional spin chain and ladder 
(La,Y,Sr,Ca)$_{14}$Cu$_{24}$O$_{41}$ 
compounds have attracted considerable interest since the
discovery of a quantum critical phase
transition in their phase diagram.\cite{Uehara} 
These compounds are the first superconducting copper oxide materials with a
non-square lattice. 
They are composed of alternately stacked chain 
and ladder planes. 
The chains are made up of CuO$_2$ linear edge-sharing CuO$_4$ squares and 
the ladder planes consist of two zig-zag strings of corner-sharing Cu$_2$O$_3$ squares
(see Fig. \ref{fig1}). 
The parent compound Sr$_{14}$Cu$_{24}$O$_{41}$ is
naturally doped with six holes per formula unit ($fu$).

Determining exactly how the holes are distributed in the system 
is hindered by the complexity of the crystal structure of Sr$_{14}$Cu$_{24}$O$_{41}$
and electron correlation effects.\cite{Vuletic,Ivek}   
Room temperature optical conductivity,\cite{Osafune} 
Near-Edge X-ray Absorption Fine Structure (NEXAFS),\cite{Nucker} 
X-ray emission spectroscopy,\cite{Kabasawa} 
and Hall coefficient measurements\cite{Tafra} 
on Sr$_{14}$Cu$_{24}$O$_{41}$ 
estimated that at least $five$ holes per $fu$ reside in chains and at most $one$ in ladders.  
High density of holes in chains is compatible with the low temperature 
T$_{C}$ $\approx$ 200~K\cite{Takigawa} 
charge ordering with the 5-fold $chain$ periodicity accompanying the 
antiferromagnetic (AF) spin-dimerization, 
observed by inelastic neutron scattering.\cite{Matsuda,Eccleston,Regnault}  
In ladders, a gapped spin-liquid and the charge density wave (CDW) appear in the ground state, 
in the spin and charge sectors, respectively. 
Whereas the spin-liquid and its collective spin excitations (triplons) are well understood 
theoretically and experimentally,\cite{Troyer,Nothbohm,Schlappa} the low hole density in the charge 
sector is at variance with the 5-fold periodicity in $ladder$ cell units (Wigner hole crystal)  
observed by low temperature resonant X-ray diffraction measurements at the O K edge if a 
4k$_F$ CDW picture is assumed.\cite{Abbamonte} 
On the other hand, a revisited interpretation of NEXAFS spectra\cite{Rushidi} 
claimed a distribution of 2.8 holes in ladders and 3.2 holes in chains, which satisfactorly explains 
the CDW in ladders, while it fails to explain AF dimer order in chains.

The apparently contradictory results are
mainly related to the lack of a suitable theoretical model. 
All published NEXAFS analyses are based on semi-empirical models that do not
take explicitly into account the complex interplay between 
spins and holes in ladders and chains.
Thus the basic question of where the holes reside in
Sr$_{14}$Cu$_{24}$O$_{41}$ has remained unresolved up to now.  

The hole distribution between chains and ladders is the key to stabilization 
of interdependent electronic phases in these two subsystems. 
It is a prerequisite to understanding the low-energy physics of the system 
and such open issues as the occurrence of Zhang-Rice singlets,\cite{ZRS} 
the evolution of spin order and the origin of the superconductivity in the Ca-doped
compound.\cite{Vuletic} 
We bridge the gap between theory and experiment by performing 
low temperature electron yield polarization dependent 
O K pre-edge NEXAFS measurements and state-of-the-art electronic structure calculations.
In the theoretical modeling we consider the full 316-atom AF unit cell,
include the core-hole effects, core level shift and
correlations in a DFT+U framework.
This method has proven to be fairly successful in reproducing 
X absorption spectra in correlated metals with well localized orbitals
\cite{Gougoussis1, Anisimov}. 
We show that, contrary to previous claims,
correlation effects and AF order stabilize holes on chains, 
induce hole-depletion in ladders, where rung sites become less populated 
than leg sites. 

\section{\label{sec:Exp}Experimental and calculation details}
High-quality single crystals of Sr$_{14}$Cu$_{24}$O$_{41}$ were grown by the 
traveling-solvent floating-zone method 
and characterized by X-ray diffraction measurements. 
A well oriented sample was cleaved $in$-$situ$ along the 
(\textbf{a},\textbf{c}) plane under a pressure of 10$^{-10}$~mbar.
The O K edge NEXAFS measurements 
were performed using the helical undulator Dragon beam line ID08 
at the European Synchrotron Radiation Facility (ESRF) 
in the total electron yield mode with 200~meV resolution. 
The incident light was normal to the sample surface and its polarization 
was parallel to the \textbf{a}/\textbf{c} sample axis. 

NEXAFS calculations are performed by the XSPECTRA\cite{Gougoussis1,Gougoussis2,Taillefumier} code 
based on Density Functional Theory (DFT)\cite{Giannozzi} and all-electron wave function 
reconstruction.\cite{Bloechl}  
Correlation effects are simulated in a mean field DFT+U framework 
\cite{Anisimov,Cococcioni} using a Hubbard on-site energy of U$_{dd}=10$~eV for copper and 
U$_{pp}= 4$~eV for oxygen.
The chain and ladder atomic positions were taken from Gotoh et al.\cite{Gotoh} 
without imposing any superstructure modulations, 
either along the chains or along the ladders.
We performed a specific NEXAFS calculation 
for $each$ of the five inequivalent oxygen sites existing in the structure, 
two in the ladders, and three in the chains. 
We find a negligible  core-level-shift.
\begin{figure}[t]
\begin{center}
\includegraphics [width=9cm,angle=0]{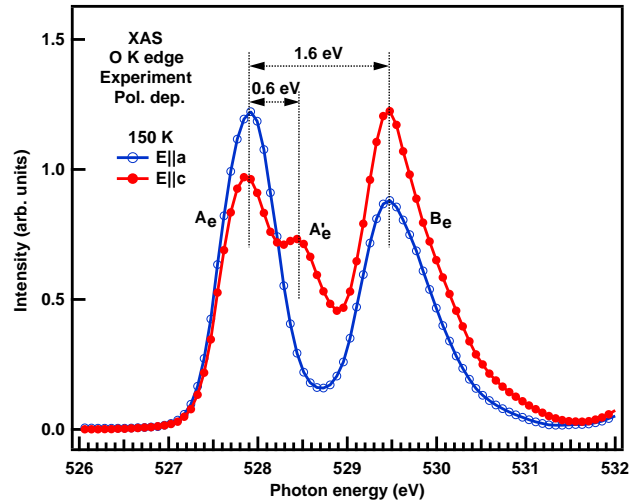}
\caption{\label{fig2} (Color online) 
Polarization dependent O K NEXAFS pre-edge intensity for 
\textbf{E}$\parallel$\textbf{a} and \textbf{E}$\parallel$\textbf{c} measured at 150 K.). 
Subscript $e$ is for experiment.}
\end{center}
\end{figure}
\begin{figure}[t]
\begin{center}
\includegraphics [width=9cm,angle=0]{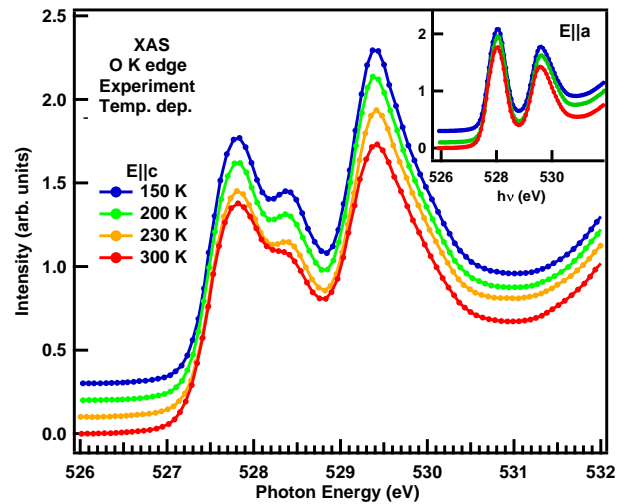}
\caption{\label{fig3} (Color online) 
Temperature dependence of the O K NEXAFS pre-edge intensity for 
\textbf{E}$\parallel$\textbf{c} showing that for this polarization spectra change 
below and above the charge ordering temperature ($\approx$ 200 K). 
The inset shows that the \textbf{E}$\parallel$\textbf{a} spectra 
temperature change is negligible.}
\end{center}
\end{figure}

\section{\label{sec:Res}Results and Discussion}

\subsection{\label{sec:NEXAFS}NEXAFS measurements}
The low temperature (150~K, see Fig.~\ref{fig2}) 
O K NEXAFS pre-edge spectra has two distinct structures, 
A$_e$, centered at 527.9~eV and B$_e$, at 529.5~eV ($e$ is for ``experiment''),
in agreement with previously published room-temperature spectra.\cite{Nucker, Rushidi, Kabasawa}  
The shape of the two structures is dependent on the polarization of the incident photon: 
for \textbf{E}$\parallel$\textbf{a}, A$_e$ is stronger than B$_e$, 
while for \textbf{E}$\parallel$\textbf{c} (direction along the chains
and the ladders) it is the opposite. 
For \textbf{E}$\parallel$\textbf{c} structure A$_e$ has a second, 
well resolved peak at 528.5~eV, A'$_e$. The intensity
of this structure is only weakly affected by the charge ordering
transition (see Fig.~\ref{fig3}). 
The \textbf{E}$\parallel$\textbf{a} spectra are even less affected at T$_{C}$  
(see Fig.~\ref{fig3} inset). 
The attribution of the features in these spectra has so far been very controversial.\cite{Nucker, Rushidi, Abbamonte} 

\subsection{\label{sec:NEXcalc}Calculation of the NEXAFS spectra}
\begin{figure}[t!]
\begin{center}
\includegraphics [width=9cm,angle=0]{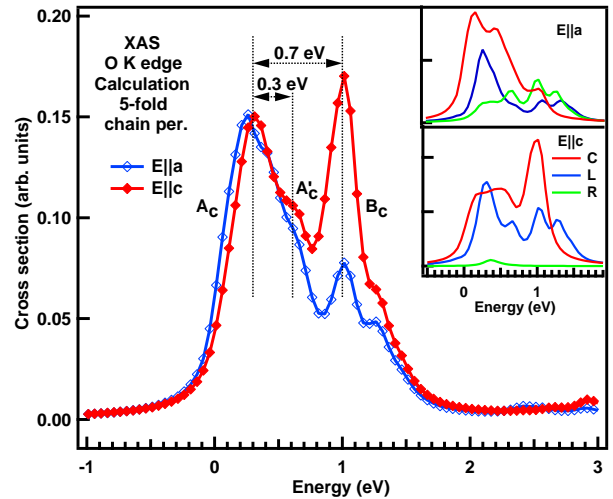}
\caption{\label{fig4} (Color online) 
Calculated O K NEXAFS pre-edge cross section on the AF unit cell with 5 fold 
periodicity in chains, for 
\textbf{E}$\parallel$\textbf{a} and \textbf{E}$\parallel$\textbf{c}. 
Insets : contribution of chain (C), ladder-leg (L) and ladder-rung (R) oxygens 
for \textbf{E}$\parallel$\textbf{a} (upper inset) and 
\textbf{E}$\parallel$\textbf{c} (lower inset). 
Zero energy corresponds to a photon energy of 527.4~eV. 
Subscript $c$ stands for calculated.}
\end{center}
\end{figure}

Theoretical modeling using the AF unit cell 
with 5-fold periodicity in chains is shown in Fig.~\ref{fig4}. 
The two, A$_e$ and B$_e$, structures are well reproduced in the calculations  
(labeled A$_c$ and B$_c$ where $c$ is for the calculated spectra),
although their separation is only 0.7~eV, compared to 
1.6 eV for the experiment. This discrepancy is attributed to the 
well-known underestimation of the Hubbard gap in DFT+U simulations.
The agreement between theory and experiment is particularly good for the 
\textbf{E}$\parallel$\textbf{c} polarization as in the \textbf{c}-direction the charges 
are less localized. 
For \textbf{E}$\parallel$\textbf{c}  
(i) the B$_c$ cross section is stronger than the A$_c$  
cross section, and (ii) the high-energy shoulder A'$_c$ is clearly visible.
For \textbf{E}$\parallel$\textbf{a}  polarization, the B$_c$ structure has smaller 
cross section than the A$_c$ peak.

The two insets in Fig.~\ref{fig4} show the calculated cross section contribution of 
chain (C), ladder-leg (L) and ladder-rung (R) oxygens to the NEXAFS
spectra.  The contribution from ladder-rungs is small or negligible
for all features except for the
B$_c$ peak in the  \textbf{E}$\parallel$\textbf{a} geometry. 
This result is in line with the analysis by N\"{u}cker et al.\cite{Nucker} and is in 
strong disagreement with the results of 
Rusydi et al.\cite{Rushidi} where the contribution of ladder rungs was over-estimated. 
Our results demonstrate that {\it ladder-rungs are hole-depleted in} 
Sr$_{14}$Cu$_{24}$O$_{41}$. 

For both polarizations the low energy A$_c$ peak is equally composed of chain 
and ladder-leg states, with a minor contribution from ladder-rungs in 
the \textbf{E}$\parallel$\textbf{a} geometry. 
Its high energy A'$_c$ shoulder is composed of both 
ladder-leg and $chain$ contributions.
This result is in stark disagreement with all previous claims
based on semi-empirical NEXAFS analysis \cite{Rushidi,Nucker} 
that attribute the A'$_c$ structure solely to ladders. 
It points to the fact that the O K edge resonant diffraction results, 
performed at the photon energy corresponding to A'c, 
could be related to the modulation reflections due to the chain-ladder 
lattice mismatch\cite{Zimmermann}. 
The Wigner-hole crystallization analysis of the O K edge resonant 
soft X-ray scattering data in Ref. \onlinecite{Abbamonte} now becomes questionable.

\begin{figure}[t!]
\begin{center}
\includegraphics [width=9cm,angle=0]{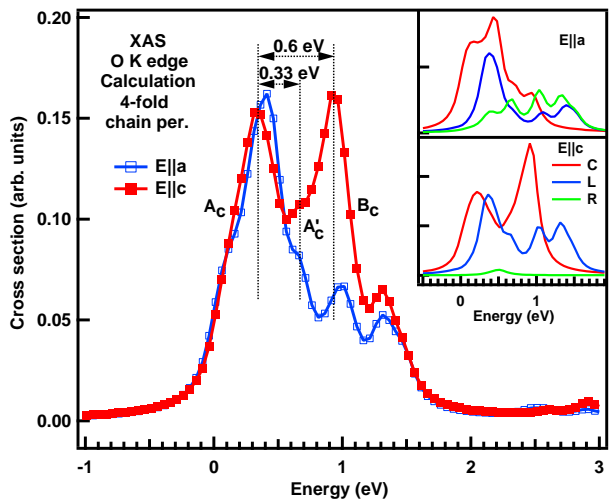}
\caption{\label{fig5} (Color online) 
Calculated O K NEXAFS pre-edge cross section on the AF unit cell with 
4 fold periodicity in chains, for 
\textbf{E}$\parallel$\textbf{a} and \textbf{E}$\parallel$\textbf{c}. 
Insets : contribution of chain (C), ladder-leg (L) and ladder-rung (R) oxygens 
for \textbf{E}$\parallel$\textbf{a} (upper inset) and 
\textbf{E}$\parallel$\textbf{c} (lower inset). 
Zero energy corresponds to a photon energy of 527.4~eV.}
\end{center}
\end{figure}
Calculations using the AF unit cell with 4-fold periodicity in chains\cite{Cox} 
is shown in Fig.~\ref{fig5}. 
The two insets show the calculated cross section contribution of 
chain (C), ladder-leg (L) and ladder-rung (R) oxygens, equivalent to these in 
Fig.~\ref{fig4}.
The experimental results are less well reproduced compared 
to the 5-fold chain-periodicity unit cell. 
First, the A$_c$ - B$_c$ energy separation is smaller. 
Further, for the \textbf{E}$\parallel$\textbf{c} geometry A$_c$ and B$_c$ are 
almost equal in intensity, contrary to the experiment. 
Finally, A'$_c$ appears as a low energy shoulder to structure B$_c$ 
and not as a high energy shoulder to A$_c$. 
For these reasons, the 4-fold chain-periodicity unit cell can be eliminated as a 
model for chain spin ordering.
\begin{figure}[h]
\begin{center}
\includegraphics [width=8cm,angle=0]{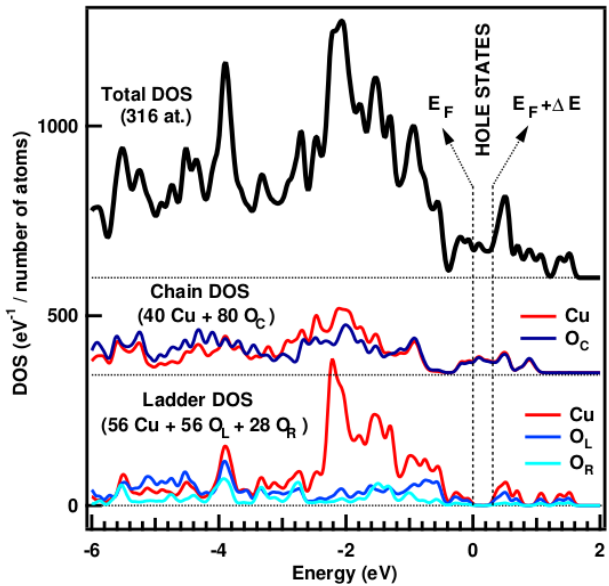}
\caption{\label{fig6} (Color online) 
(upper panel) Total DOS of Sr$_{14}$Cu$_{24}$O$_{41}$ without a core-hole, 
calculated on the AF unit cell with 5 fold chain periodicity. 
The Fermi level (E$_F$) corresponds to 0~eV. 
(lower panels) Local chain- and ladder-DOS projected to Cu (red line) and O (blue lines). 
The second dotted line indicates where E$_F$ would be if the system had 6 
additional electrons per $fu$, i.e. no holes (E$_F$' = E$_F$+$\Delta E$, 
where $\Delta E$ = 0.31~eV corresponds to additional 6 x 4 = 24 electrons).}
\end{center}
\end{figure}
\begin{figure}[h]
\begin{center}
\includegraphics [width=8.5cm,angle=0]{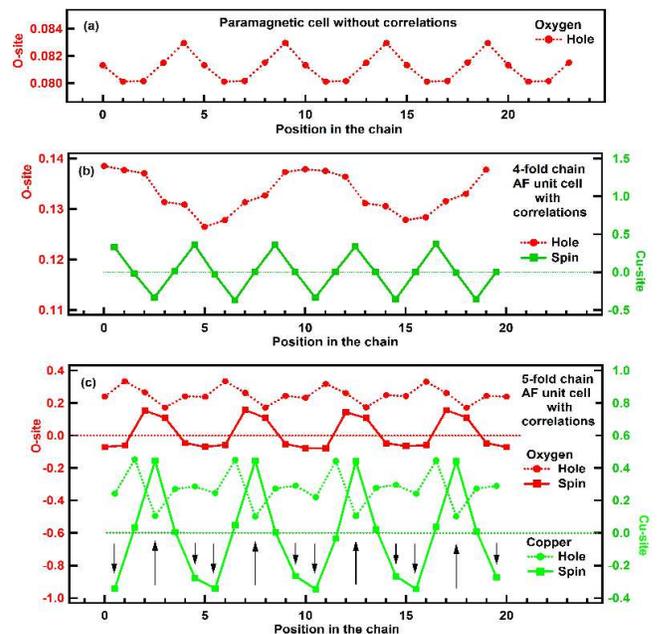}
\caption{\label{fig7} (Color online) 
Chain site dependent hole (spin) density :
(a) Chains have natural 5-fold hole periodicity in the paramagnetic cell without correlations ; 
(b) When the 4-fold chain spin periodicity is imposed, the 5-fold hole periodicity is destroyed ; 
(c) In the case of the 5-fold spin periodicity, the natural hole periodicity is preserved, 
on oxygen and as well on copper sites.}
\end{center}
\end{figure}
\subsection{\label{sec:DOScalc}Density of states}
\begin{figure}[h]
\begin{center}
\includegraphics [width=6cm,angle=-90]{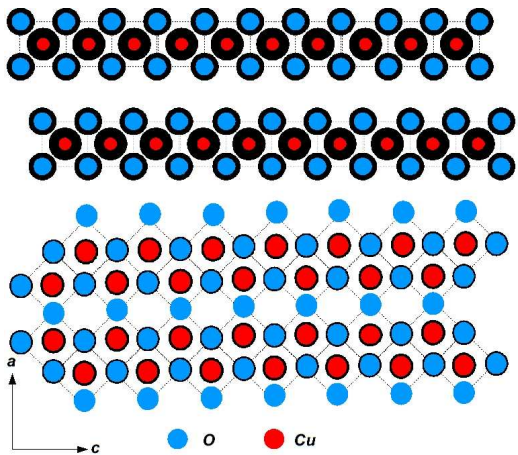}
\caption{\label{fig8} (Color online) 
The thickness of the contour at O/Cu sites symbolizes the mean hole density per site 
in chains (upper panel) and ladders (lower panel) as determined by $ab$-$inition$ calculations.}
\end{center}
\end{figure}

To understand the role of core-hole effects we also performed density of states (DOS) calculations 
for the system without a core-hole (see Fig.~\ref{fig4}). 
Contrary to previous  theoretical works,\cite{Arai,Schwing,Ma} performed on a minimal cell
without AF ordering, we find the occurrence of a  
0.25~eV gap appearing in ladder states. This finding is
in agreement with the insulating behavior of the ladders 
\cite{Motoyama,Blumberg,Vuletic2005}  and points out the need of including the
full AF crystal structure in the simulation. 
The occurrence of a 0.25~eV Hubbard gap in ladder states results in a strong hole depletion of this sub-system. 
Indeed, we find that only about 0.4 holes reside in ladders, while the chains 
accommodate about 5.3 holes per $fu$. 
A more detailed analysis of  hole distribution is presented in Table~\ref{proba}.
The mean value of the probability of accommodating a hole on one atom is schematically presented 
in Fig.~\ref{fig8} by the thickness of the contour surrounding atoms of each group (chain, ladder-leg, 
ladder-rung).

\begin{table}[h!]
\begin{center}
\begin{tabular}{c|ccc|rcc|c}
\hline
\hline
& \multicolumn{3}{|c|}{Chain-}&\multicolumn{3}{|c|}{Ladder-}&\multicolumn{1}{|c}{Sr-inter-layer}\\
\hline
$n$& \multicolumn{3}{|c|}{5.3}&\multicolumn{3}{|c|}{0.4}&\multicolumn{1}{|c}{0.3}\\
\hline
atom&Cu&&O&Cu&O(leg)&O(rung)&Sr\\
$N_g$&10&&20&14&14&7&14\\
$n_{g}$&2.75&&2.55&0.21&0.14&0.05&0.3\\
\hline
\hline
\end{tabular}
\end{center}
\caption{\label{proba} Calculated fraction ($n$) of 6 holes per $fu$ 
residing in chain-, ladder- and in Sr-inter-layers
of Sr$_{14}$Cu$_{24}$O$_{41}$; 
number of atoms in each group ($N_g$) ;
distribution of holes between Cu- and O-sites in chains and ladders ($n_g$).}
\end{table}

The role of correlations is even more evident when analyzing hole and
spin order along the chains in the paramagnetic DFT calculations and
in the DFT including U and AF ordering. When correlation
effects are switched-on the number of holes in the ladder is decreased by
a factor of five.
{\it Hole-depletion of ladders is thus a pure correlation effect}. 

The distribution of more than 5 holes in chains and less 
than 1 in ladders is compatible not only with the AF unit cell with 5-fold chain periodicity, 
but also with the AF model with 4-fold chain periodicity.\cite{Cox} 
But the latter model turns out to be irrelevant since in this case the NEXAFS calculations 
could not reproduce the experimental data (see Sect.\ref{sec:NEXcalc}). 
Moreover, the 4-fold chain-spin periodicity destroys the natural 5-fold hole ordering, while the 
5-fold spin periodicity preserves it, as shown in Fig.\ref{fig7}.
We would like to stress once more that taking into account the proper long-range AF order
is crucial to achieve a correct description of O K-edge NEXAFS spectra.

It is worthwhile noting that our DOS is in good agreement with
experiments as we find the occurrence of a strong -2~eV band with mixed
chain/ladder contributions and large Cu-ladder weight, 
in accordance with Angle Resolved Photoemission Spectroscopy (ARPES) 
measurements.\cite{Takahashi} 
Furthermore the two ladder-DOS structures at -0.25~eV and -0.1~eV have also been   
observed in a recent ARPES \cite{Yoshida} experiment and were identified as the quasi-1D underlying 
Fermi surface.

Finally, the tendency of hole-depleting ladder-rungs found in our work, provides a satisfactory 
explation for the 2D to 1D cross-over when the number of holes in the system is reduced,
like in under-doped compounds\cite{Ivek} or at low temperature where the 
less localized ladder sub-system suffers from the migration of holes to chains (back-transfer) 
as reported in the NMR.\cite{Piskunov}  
On the fully doped side we suggest that the Ca doping has a twofold effect: 
it destroys long-range AF order in chains \cite{Matsuda,Nagata,Isobe} and reduces hole-depletion 
in ladders. 
Increased population of the rung sites reinforces the 2D character of the ladders. 
For sufficient Ca-induced ladder-doping the CDW ground state
is then suppressed in favor of the 2D superconductivity under pressure.\cite{Uehara, VuleticPRL}

\section{\label{sec:Concl}Conclusion}
In conclusion, our experimental NEXAFS spectra and first-principles theoretical modeling including 
the complete 316 atom AF unit cell, core-hole and correlation effects 
demonstrate that holes mainly reside on chains and thus ladders are hole-depleted. 
This finding resolves long standing controversial interpetations by several 
authors based on semi-empirical models. 
The analysis by N\"{u}cker et al.\cite{Nucker} is in agreement with 
our findings based on an $ab$-$initio$ approach. 
It is clear, however, that their arguments were based on a too limited description 
since correlations and long range AF behavior consisting of a 5-fold chain periodicity 
must be included to obtain the corrrect interpretation of the experimental spectra.

In ladders, majority of holes reside in leg sites, while only a tiny minority populates rung sites. 
Further experiments and calculations are under consideration to verify our suggestion that the small 
affinity of rung sites in the parent compound explains the 1D - 2D cross-over going from the 
under-doped to the fully doped side in the phase diagram of (La,Y,Sr,Ca)$_{14}$Cu$_{24}$O$_{41}$.

Our work unambiguously answers the out-standing question of where the holes reside in  Sr$_{14}$Cu$_{24}$O$_{41}$. 
This is crucial to understanding all phenomena occurring in the 
phase diagram of this family of compounds, ranging from charge-ordered antiferromagnetism to 
superconductivity and is a prerequisite to all investigations based on correlated 
models of spin ladder and chain compounds.\cite{Vuletic}   

\section{\label{sec:Ack}Acknowledgments}
M. C. and C. G.  acknowledge discussions with F. Mauri. 
Calculations were performed at the  IDRIS superconducting center (project
number 96053). 
S. T. acknowledges support from the Croatian Ministry of Science, Education and Sports under grant 
035-0000000-2836.

%\bibitem{Zimmermann} M. v. Zimmermann, J. Geck, S. Kiele, R. Klingeler, 
%and B. B\"{u}chner, Phys. Rev. B \textbf{73} 115121 (2006)
%
%\textbf{}

\end{document}